# Subdiffractional focusing and guiding of polaritonic rays in a natural hyperbolic material

S. Dai[1], Q. Ma[2], T. Andersen[2], A. S. Mcleod[1], Z. Fei[1], M. K. Liu[1,3], M. Wagner[1], K. Watanabe[4], T. Taniguchi[4], M. Thiemens[5], F. Keilmann[6], P. Jarillo-Herrero[2], M. M. Fogler[1], D. N. Basov[1]*.

[1]*Department of Physics, University of California, San Diego, La Jolla, California 92093, USA*

[2]*Department of Physics, Massachusetts Institute of Technology, Cambridge, Massachusetts 02215, USA*

[3]*Department of Physics, Stony Brook University, Stony Brook, New York 11794, USA*

[4]*National Institute for Materials Science, Namiki 1-1, Tsukuba, Ibaraki 305-0044, Japan*

[5]*Department of Chemistry and Biochemistry, University of California, San Diego, La Jolla, California 92093, USA*

[6]*Ludwig-Maximilians-Universität and Center for Nanoscience, 80539 München, Germany*

*Correspondence to: dbasov@physics.ucsd.edu

**Abstract**

Uniaxial materials whose axial and tangential permittivities have opposite signs are referred to as indefinite or hyperbolic media. In such materials light propagation is unusual, leading to novel and often non-intuitive optical phenomena. Here we report infrared nano-imaging experiments demonstrating that crystals of hexagonal boron nitride, a natural mid-infrared hyperbolic material, can act as a "hyper-focusing lens" and as a multi-mode waveguide. The lensing is manifested by subdiffractional focusing of phonon-polaritons launched by metallic disks underneath the hexagonal boron nitride crystal. The waveguiding is revealed through the modal analysis of the periodic patterns observed around such launchers and near the sample edges. Our work opens new opportunities for anisotropic layered insulators in infrared nanophotonics complementing and potentially surpassing concurrent artificial hyperbolic materials with lower losses and higher optical localization.

**Introduction**

One of the primary goals of nanophotonics is concentration of light on scales shorter than the free-space wavelength $\lambda$. According to the general principles of Fourier optics, this is only possible provided electromagnetic modes of large tangential momenta $k_t > \omega/(2\pi)$, normally evanescent, are nonetheless able to reach the focal plane (the $x$-$y$ plane). Here $\omega = \lambda^{-1}$ is the measure of frequency common in spectroscopy and $k_t = \sqrt{k_x^2 + k_y^2}$. In devices known as superlenses[1-6], this requirement is realized via resonant tunneling between the opposite sides of the structure. However, the tunneling is very sensitive to damping, e.g., the magnitude of the imaginary part of the permittivity $\varepsilon$ of the superlens material[7]. The largest characteristic momentum that can pass through a superlens of thickness $d$ can be found from the relation $\operatorname{Im} \varepsilon \sim e^{-k_t d}$. In this regard, hyperbolic media (HM)[8,9] promise a significant advantage as they support large-$k$ hyperbolic polaritons that remain propagating rather than evanescent, so that the condition on damping is much softer (see below). The unusual properties of hyperbolic polaritons in HM[8-20] stem from the dispersion of these modes that is described by the equation of a hyperboloid:

$$\varepsilon_t^{-1} k_z^2 + \varepsilon_z^{-1}(k_x^2 + k_y^2) = (2\pi\omega)^2, \qquad (1)$$

where $\varepsilon_z$ and $\varepsilon_t \equiv \varepsilon_x = \varepsilon_y$ are the axial and tangential permittivities, respectively. The hyperboloid is single-sheeted if $\varepsilon_z > 0$, $\varepsilon_t < 0$ (Type II) and two-sheeted if $\varepsilon_z < 0$, $\varepsilon_t > 0$ (Type I), see Figs. 1a and 1b, respectively. In both cases the slope of the propagation (group velocity) direction, which is orthogonal to the dispersion surface, asymptotically approaches

$$\tan\theta(\omega) = i\frac{\sqrt{\varepsilon_t(\omega)}}{\sqrt{\varepsilon_z(\omega)}}. \qquad (2)$$

The condition for achieving super-resolution is $\text{Im } k_z d = (k_t d)\text{Im tan }\theta \sim 1$. Hence, admissible $\text{Im }\varepsilon_z, \text{Im }\varepsilon_t$ scale algebraically rather than exponentially with the resolution $k_t^{-1}$.

Directional propagation of hyperbolic polaritons along "resonance cones" of apex angle $\theta$ has been observed in a magnetized plasma[21,22], which behaves as a natural HM in the microwave domain. A major resurgence of interest to HM was prompted by their discussion in the context of artificial materials (metamaterials)[23,24]. Examples of such hyperbolic metamaterials include microstrip arrays, where directional propagation and focusing of hyperbolic polaritons have been experimentally observed[25,26]. Directional optical beams have been studied in planar[25-28] and curved[12,29] metamaterials made of alternating layers of metals and semiconductors. The work on non-planar structures[12,29] was motivated by theoretical proposals of a hyperlens[30-32], a device in which directional beams outgoing from a subdiffractional source enable optical magnification. However, improvement over the diffraction limit has so far been severely impeded by losses in constituent metals and imperfections of nanofabrication.

Recent work[33,34] identified hexagonal boron nitride (hBN) as a low-loss natural HM in the mid infrared (IR) domain. This layered insulator has emerged as a premier substrate or a spacer for van der Waals heterostructures[35,36]. Light atomic masses, strong anisotropy, and the polar band between B and N yield prominent optical phonon modes that create two widely separated stop-bands – spectral intervals where one of the principal values of the dielectric tensor is negative[33,34,37]. The upper band comprises

$\omega = 1370 - 1610$ cm$^{-1}$ where the real part of $\varepsilon_t$ (the in-plane permittivity) is negative while that of $\varepsilon_z$ is positive. In the lower band spanning $\omega = 746 - 819$ cm$^{-1}$, the signs of the permittivity components are reversed. Thus, the out-of-plane crystal vibrations enable the Type I hyperbolic response, whereas the in-plane ones accounts for the Type II behavior. The momentum-frequency dispersion surface for the hyperbolic polaritons of the upper band resembles a "butterfly" (Fig. 1c) composed of individual hyperbolas sketched in Fig. 1a. It can be contrasted with the flat dispersion surfaces of longitudinal phonons typical for isotropic materials. Effectively, in hBN the longitudinal phonons are hybridized with the transverse ones by quasi-static Coulomb interaction mediated by large-$k$ photons[38]. Because the hyperbolic response in hBN originates from the anisotropic phonons, in the following, the large-$k$ hyperbolic polaritons are referred to as hyperbolic phonon polaritons (HP$^2$).

**Results**

*Subdiffractional focusing and imaging through hBN*

In our experiments, efficient excitation and detection of HP$^2$ in hBN are accomplished with the help of optical antenna structures[39,40]. The antennas concentrate electric field and bridge the large momentum mismatch between the free-space photons and the HP$^2$. In our previous work[33], we used for this purpose a sharp tip of an atomic force microscope (AFM) incorporated in our scattering-type scanning near-field optical microscopy (s-SNOM) apparatus (Methods). Here we additionally demonstrate the antenna and polariton-launching capabilities of Au disks patterned on a SiO$_2$ substrate.

The AFM topography image in Fig. 2a depicts Au disks of diameters (top to bottom) 1000, 500 and 200 nm and thickness of about 50 nm. After the subsequent deposition of hBN crystals of thickness $d = 100 – 1060$ nm and lateral sizes up to 10 µm, these Au disks become encapsulated between hBN and $SiO_2$. The hBN crystal remains essentially flat, as verified by AFM. Below we present experimental results demonstrating that interaction of these disks with an incident IR beam excites polaritons that travel across hBN and produce specific contrast patterns at the other surface. We show that the observed dependence of the near-field images on the frequency and hBN thickness is the result of directional propagation of the polaritons along conical surfaces with frequency-tunable apex angle given by equation (2). Thus, hBN may emerge as a new standard bearer for mid-IR nanophotonics by enabling devices for deeply subdiffractional propagation, focusing, and imaging with tunable characteristics.

Representative s-SNOM imaging data are shown in Fig. 2. Figure 2b depicts an s-SNOM scan taken at the top surface of hBN of thickness $d = 395$ nm at frequency $\omega = 1515$ cm$^{-1}$ ($\lambda = 6.6$ µm). Here we plot the third harmonics of the scattering amplitude $s(\omega)$ (Methods). In this image each Au disk is surrounded by a series of concentric "hot rings" of strongly enhanced nano-IR contrast. The diameters of all the disks are much smaller than $\lambda$ (see also Fig. 2a), the smallest one being 200 nm = $\lambda/33$. The diameters of the hot rings can be larger, smaller, or equal to those of the disks. The spacing between adjacent hot rings in the same sample increases with the IR frequency but decreases with the sample thickness. We stress that images displayed in Fig. 2b could only be detected if the IR wavelength falls inside the hyperbolic spectral regions. Outside of the hBN stop bands, no hot rings can be identified by the s-SNOM. In fact, the entire image is homogeneous, comprised of nothing but random noise, as illustrated by Fig. 2d for $\omega = 1740$ cm$^{-1}$ ($\lambda =$ 5.7 μm).

We now elaborate on the formation of images in Fig. 2 recorded with our s-SNOM apparatus with the help of a model of HP$^2$ propagation through a slab of hBN (Figs. 3a, c and d). Consider a perfectly thin metallic disk sandwiched between a slab of a HM of thickness $d$ and an isotropic dielectric substrate. The system is subject to a uniform electric field of frequency ω and amplitude $E_0$ in the x-direction. An approximate solution for the total field in this system can be found analytically (Supplementary Note 1). The corresponding distributions of the z-component of the field $E_z(x, y, z)$ in the two cross-sections, $y = 0$ (the vertical symmetry plane) and $z = d - 0$ (just below the top surface of the hBN slab), are illustrated in Fig. 3c. These plots are computed for three representative radii of the disk using permittivity values at $\omega = 1515$ cm$^{-1}$. The plots demonstrate a series of concentric high-intensity rings on the top surface, very similar to the data in Fig. 2b. The interpretation (Figs. 3a, c) is straightforward: the external field polarizes the disk, which perturbs the adjacent HM (hBN in our case) and launches polaritons. The HP$^2$ emission occurs predominantly at disk edges due to the high concentration of electric field therein. Polaritonic rays propagate across the slab, maintaining a fixed angle $\theta$ with respect to the z-axis: the "resonance cone" direction[18,21,22, 25-28]. Upon reaching the other slab surface, they undergo a total internal reflection with the reflected cone extending toward the bottom surface. The process repeats until eventually the field vanishes because of radial spreading and/or damping. The role of the s-SNOM tip in imaging experiments in Fig. 2 is to out-couple HP$^2$ fields at the top surface (Fig. 3a). The observed s-SNOM signal is roughly proportional to the amplitude of the electric field immediately above the slab $E_z(z = d + 0)$. (Note that it is related to the field just inside the slab by a constant factor, $E_z(z = d + 0) = \varepsilon_z(\omega) E_z(z = d - 0)$.)

The above model of image formations via HP$^2$ yields a number of quantitative predictions that are in accord with our observations. The scenario of oblique propagation implies that upon each roundtrip across the slab, the excitation front returns to the same surface displaced radially by the distance

$$\delta = 2 \tan \theta(\omega) d . \tag{3}$$

Accordingly, the radii of the "hot rings" at the top surface of the slab are given by

$$r_n = \left| a + \left(n - \tfrac{1}{2}\right) |\delta| \right|, \quad n = 0, \pm 1, \pm 2, \ldots \tag{4}$$

where $a$ is the disk radius. The intensity of the rings is expected to decrease with $|n|$. Consistent with this formula, the smallest rings in Fig. 3c have the radius $r_0 = |a - |\delta|/2|$. Particularly interesting is the case where the innermost ring shrinks to a single bright spot, $r_0 = 0$. Experimentally, we observed spots of diameter 200 nm (the full width at half maximum, see Supplementary Note 1), which corresponds to $\lambda/33$ for Fig. 2b (top). Focal spots of similar size 185 – 210 nm were observed in all other hBN crystals, with the thickness up to 1050 nm (Supplementary Figure 5).

A proposal for focusing of electromagnetic radiation via resonance-cone propagation in hyperbolic media was theoretically discussed in the context of magneto-plasmas[21]. Experimental confirmation of this idea in an artificial hyperbolic multi-layer was reported where $\lambda/6$ focusing was deduced from examining the pattern of a polymerized photoresist behind a two-slit polaritonic launcher[26]. Here, using a natural hyperbolic slab (hBN crystal), we demonstrated the $\lambda/33$ focusing in both spatial directions via out-coupling of polaritons with the IR nano-probe. We stress that a distinction should be made between 'focusing' and 'imaging.' Focusing devices can be of both imaging and non-imaging type[41] and both are important in applications. Our hBN device (Fig. 3a) is an example of the latter.

Continuing with the verifiable predictions of our model, we note that equations (3) and (4) indicate that the slope $\tan\theta$ of the resonance cone is uniquely related to the radii of the hot rings (Fig. 3a). To test this prediction we analyzed images collected from samples of different hBN thicknesses and different Au disk diameters. For each of these, we determined the radius $r_1$ of the 1$^{st}$-order ring and computed $|\tan\theta| = (r_1 - a)/d$ as a function of the IR frequency (Fig. 3a). As shown in Fig. 3b, all the data collapse toward a single smooth curve computed from equation (2) using optical constants of hBN from ref. 33. Yet another prediction of the model: the polaritonic rays travel along the $z$-axis provided that $\varepsilon_t(\omega)$ and therefore $\theta(\omega)$ are vanishingly small. This condition is satisfied at $\omega = 1610\,\text{cm}^{-1}$ (Fig. 2c) where we observe almost 1:1 images of Au disks. Similar behavior was observed when instead of the disks more complicated metallic shapes were imaged (Supplementary Figure 5). Thus, the totality of our data establishes the notion of directional propagation of HP$^2$ in hBN over macroscopic distances with a frequency-tunable slope (Fig. 3b).

*Real-space imaging of multiple guided polaritons in hBN*

The outlined real-space picture has a counterpart in its conjugate momentum space. Mathematically, the resonance cones in the real space are coherent superpositions of an infinite number of polariton modes of a slab. Such modes are characterized by quantized momenta, $k_{z,l} = (\pi/d)(l + \alpha)$, labeled by integer index $l$[33]. Here $\alpha \sim 1$ (in general, $\omega$-dependent) quantifies the phase shift acquired at the total internal reflection from the slab surfaces. Per equation (1), the tangential momenta of these modes are also quantized,

$$k_{t,l}(\omega) \simeq \cot\theta(\omega)\, k_{z,l}(\omega) = \frac{2\pi}{\delta(\omega)}[l + \alpha(\omega)]. \qquad (5)$$

In the last step we have applied equation (3). For illustration, the dispersion curves of such guided modes in the upper stop-band of hBN of thickness 105 nm are shown in Fig. 1c, where they are overlaid on the dispersion surface of bulk hBN. The same curves are replotted as $\omega$ vs. $k_t$ in Fig. 4a. In Fig. 4b the dispersion curves of the guided modes of lower stop-band are shown. An intriguing aspect of these curves is that their slope $\partial\omega/\partial k_t$ is positive (negative) in the upper (lower) band. This sign difference is a consequence of the opposite direction of the group velocity vector for the Type I and Type II cases, cf. Fig. 1a and 1b. Central to the connection between the resonance cones in the real space and the quantized momenta in the $k$-space is that these momenta form an equidistant sequence of period $\Delta k_t = k_{t,l+1} - k_{t,l} = 2\pi/\delta$. Therefore, if several guided modes are excited simultaneously by a source, their superposition would produce beats with period $2\pi/\Delta k_t$ in real space. This is precisely the spacing $\delta$ between periodic revivals of the "hot rings" (equation (3) and Fig. 2). Thus, the multi-ring images and the existence of higher-order guided modes are complementary manifestations of the same fundamental physics. In our previous work[33] we reported nano-imaging and nano-spectroscopic study of the lowest-momentum guided mode $l = 0$ in hBN crystals. Below we present new results documenting the first observation of the higher-order (up to three) guided modes in such materials by direct nano-IR imaging.

In order to map the dispersion of $HP^2$ we utilized hBN crystals on $SiO_2$ substrate without any intervening metallic disks (Methods). Here the sharp tip of the s-SNOM serves as both the emitter and the detector of the polariton waves on the open surface of the hBN. As the tip is scanned toward the sample edge, distinct variations in the detected scattering amplitude $s(\omega)$ are observed. Such variations are caused by passing over minima and maxima of the standing waves created by interference of the polaritons launched by the tip and their reflections off the sample edges (Fig. 5a). Representative data for the upper stop-band (the Type II hyperbolic region) are shown in Figs. 5b-f, where we plot $s(\omega)$ at various IR frequencies. Specifically, the image presented in Fig. 5b exhibits oscillations with the period ~1μm extending parallel to the edge of a 31-nm-thick hBN crystal. While these oscillations are similar to those reported previously[33], a high-resolution scan performed very close to the edge (the olive square) reveals additional oscillations occurring on a considerably shorter scale: down to hundreds of nm (Figs. 5c-e). Similar results have been obtained using many other samples. For example, Fig. 5f also shows short-scale oscillations near the edges co-existing with longer-range oscillations further away from the edge in the data collected for a thicker hBN crystal ($d = 105$ nm).

To analyze the harmonic content of the measured $s(\omega)$ quantitatively we employed the spatial Fourier transform (FT). An example shown in Fig. 5h is the FT of the line trace α from Fig. 5g. The three dominant peaks in the FT are marked with β' (blue), γ' (magenta) and ζ' (olive). These peaks have been deemed statistically significant and their positions $k_\beta$, $k_\gamma$ and $k_\zeta$ have been recorded for each of the traces studied. We reasoned that including additional weaker peaks into consideration may be unwarranted at this stage.

Indeed, the gross features in the real-space trace α exceeding the noise level of ~ 1 a.u. are accounted for by oscillations in the three partial traces β, γ and ζ, which are obtained by the inverse FT of the shaded regions in Fig. 5h.

The remaining step in the analysis is to establish the connection of thus determined momenta $k_β$, $k_γ$ and $k_ζ$ and the momenta $k_{t,l}$ of the guided modes, equation (5). This requires more care than in prior studies of single-mode waves in 2D materials[33,42-44]. The interference patterns near the edge can be created by various combinations of the tip-launched waves (labeled by $l$) and edge-reflected waves (labeled by $r$). The total momentum of a particular combination is $k_{t,l} + k_{t,r}$. If the mode index is conserved, $l = r$, the set of possible periods narrows down to $2k_{t,l}$. This is consistent with our data obtained for several IR frequencies (Figs. 4a), where the symbols indicate $k_β$, $k_γ$ and $k_ζ$. These data are in a quantitative agreement with the calculated dispersion curves for the $l$ = 0, 1 and 2 polariton guided waves in the upper stop-band. The analysis of polariton propagation length[33] shows that the loss factor is as low as $γ$ ~ 0.03 (Supplementary Figure 4). Dispersion mapping in the lower band (746 – 819 cm$^{-1}$) where no monochromatic lasers are available is discussed in Supplementary Figure 3. Broad-band lasers used in an independent study by Li et al. have allowed to demonstrate focusing behavior of hBN in this challenging frequency region[45].

**Discussion**

Data presented in Figs. 2-5 demonstrate launching, long-distance waveguiding transport, and focusing of electromagnetic energy in thin crystals of hBN. These phenomena are enabled by directional propagation of large-momentum polariton beams in this natural hyperbolic material. The sharpness of the attained focusing, $\lambda/33$ at distances up to $\lambda/6$ (Supplementary Figure 5), in units of the free-space wavelength, surpasses all prior realizations of superlenses and hyperlenses. Remarkably, a simple addition of a circular metallic launcher transforms an hBN crystal into a powerful focusing[19] device! The analysis presented in Supplementary Note 1 (Supplementary Equation 10) indicates that the size of the focal spot in our system is limited by the finite thickness ~50 nm of Au disks. By using thinner disks, say 20 nm thick, one should be able to achieve focal spots as small as ~ $\lambda/10^2$, comparable to the spatial resolution of our nano-IR apparatus. A fundamental advantage of using natural rather than artificial hyperbolic materials is the magnitude of the upper momentum cutoff. In a natural material such as hBN this cutoff is ultimately set by interatomic spacing thus immensely enhancing the spatial resolution. Additionally, we have shown that hBN can serve as a multi-mode waveguide for polaritons with excellent figure of merit: loss factor as small as $\gamma \sim 0.03$. These characteristics exceed the benchmarks[46-48] of current metal-based plasmonics and metamaterials. The physics behind this fundamental advantage of phonon polaritons over plasmons in conducting media is in the absence of electronic losses in insulators. Applications of hBN for non-imaging focusing devices[41], subdiffractional waveguides, and nanoresonators[34] readily suggest themselves[45]. Combining such elements together may lead to development of sophisticated nanopolaritonic circuits.

## Methods

### Experimental setup

The nano-imaging and nano-FTIR experiments described in the main text were performed at UCSD using a commercial s-SNOM (www.neaspec.com). The s-SNOM is based on a tapping-mode AFM illuminated by monochromatic quantum cascade lasers (QCLs) (www.daylightsolutions.com) and a broad-band laser source utilizing the difference frequency generation (DFG) (www.lasnix.com)[49]. Together, these lasers cover a frequency range of 700 – 2300 cm$^{-1}$ in the mid-IR. The nanoscale near-field images were registered by pseudo-heterodyne interferometric detection module with AFM tapping frequency and amplitude around 250 kHz and 60 nm, respectively. To obtain the background-free images, the s-SNOM output signal used in this work is the scattering amplitude $s(\omega)$ demodulated at the $n^{th}$ harmonics of the tapping frequency. We chose $n = 3$ in this work.

*Sample fabrication*

Silicon wafers with 300-nm-thick $SiO_2$ top layer were used as substrates for all samples. The Au patterns of various lateral shapes and 50-nm thickness were fabricated on these wafers by electron beam lithography. The hBN microcrystals of various thicknesses were exfoliated from bulk samples synthesized under high pressure[50]. Such microcrystals were subsequently mechanically transferred onto either patterned or unpatterned parts of the substrates.

**Acknowledgments:**

D.N.B. acknowledges support from DOE-BES grant DE-FG02-00ER45799 and the Gordon and Betty Moore Foundation's EPiQS initiative through Grant GBMF4533; research on polariton focusing is supported by AFOSR. Work at UCSD is supported by the Office of Naval Research, AFOSR, NASA and The University of California Office of the President. A.S.M. acknowledges support from an Office of Science Graduate

Research Fellowship from U.S. Department of Energy. P.J-H acknowledges support from AFOSR grant number FA9550-11-1-0225.

**Author contributions:**

All authors were involved in designing the research, performing the research and writing the paper.

**Competing interests:**

F.K. is one of the cofounders of Neaspec and Lasnix, producer of the s-SNOM and infrared source used in this work. All other authors declare no competing financial interests.

**References:**

1. Pendry, J. B. Negative refraction makes a perfect lens. *Phys. Rev. Lett.* **85,** 3966-3969 (2000).

2. Fang, N., Lee, H., Sun, C. & Zhang, X., Sub-Diffraction-Limited Optical Imaging with a Silver Superlens. *Science* **308,** 534-537 (2005).

3. Taubner, T., Korobkin, D., Urzhumov, Y., Shvets, G. & Hillenbrand, R. Near-field microscopy through a SiC superlens. *Science* **313,** 1595 (2006).

4. Zhang, X. & Liu, Z. Superlenses to overcome the diffraction limit. *Nature Mater.* **7,** 435-441 (2008).

5. Smolyaninov, I. I., Hung, Y. J., Davis, C. C. Magnifying Superlens in the Visible Frequency Range. *Science* **315,** 1699-1701 (2007).

6. Kehr, S. C. et al. Near-field examination of perovskite-based superlenses and superlens-enhanced probe-object coupling. *Nature Commun.* **2**, 249 (2011).

7. Smith, D. R., Schurig, D., Rosenbluth, M., Schultz, S., Ramakrishna, S. A. & Pendry, J. B. Limitations on subdiffraction imaging with a negative refractive index slab. *Appl. Phys. Lett.* **82,** 1506-1508 (2003).

8. Poddubny, A., Iorsh, I., Belov, P. & Kivshar, Y. Hyperbolic metamaterials. *Nature Photon.* **7,** 948-957 (2013).

9. Guo, Y., Newman, W., Cortes, C. L. & Jacob, Z. Applications of hyperbolic metamaterial substrates. *Advances in OptoElectronics* **2012,** 452502 (2012).

10. Hoffman, A. J. et al. Negative refraction in semiconductor metamaterials. *Nature Mater.* **6,** 946-950 (2007).

11. Yao, J. et al. Optical negative refraction in bulk metamaterials of nanowires. *Science* **321,** 930 (2008).

12. Liu, Z., Lee, H., Xiong, Y., Sun, C. & Zhang, X. Far-field optical hyperlens magnifying sub-diffraction-limited objects. *Science* **315,** 1686 (2007).

13. Yang, X., Yao, J., Rho, J., Yin, X. & Zhang, X. Experimental realization of three-dimensional indefinite cavities at the nanoscale with anomalous scaling laws, *Nature Photon.* **6,** 450-454 (2012).

14. Argyropoulos, C., Estakhri, N. M., Monticone, F. & Alu, A. Negative refraction, gain and nonlinear effects in hyperbolic metamaterials. *Opt. Express* **21,** 15037-15047 (2013).

15. Biehs, S. A., Tschikin, M. & Ben-Abdallah, P. Hyperbolic metamaterials as an analog of a blackbody in the near field. *Phys. Rev. Lett.* **109,** 104301 (2012).

16. Noginov, M. A. et al. Controlling spontaneous emission with metamaterials. *Opt. Lett.* **35**, 1863-1865 (2010).

17. Guo, Y., Cortes, C. L., Molesky, S. & Jacob, Z. Broadband super-Planckian thermal emission from hyperbolic metamaterials. *Appl. Phys. Lett.* **101,** 131106 (2012).

18. Jacob, Z., Smolyaninov, I. I. & Narimanov, E. E. Broadband Purcell effect: Radiative decay engineering with metamaterials. *Appl. Phys. Lett.* **100,** 181105 (2012).

19. Smith, D. R., Schurig, D., Mock, J. J., Kolinko, P. & Rye, P. Partial focusing of radiation by a slab of indefinite media. *Appl. Phys. Lett.* **84,** 2244-2246 (2004).

20. Vinogradov, A. P., Dorofeenko, A. V. & Nechepurenko, I. A. Analysis of plasmonic Bloch waves and band structures of 1D plasmonic photonic crystals. *Metamaterials* **4**, 181-200 (2010).

21. Fisher, R. K. & Gould, R. W. Resonance cones in the field pattern of a short antenna in anisotropic plasma. *Phys. Rev. Lett.* **22,** 1093-1095 (1969).

22. Levine, B., Greene, G. J. & Gould, R. W. Focusing resonance cones. *Phys. Fluids* **21**, 1116 -1119 (1978).

23. Lindell, I. V., Tretyakov, S. A., Nikoskinen, K. I. & Ilvonen, S. BW media-media with negative parameters, capable of supporting backward waves. *Microw. Opt. Technol. Lett.* **31**, 129-133 (2001).

24. Smith, D. R. & Schurig, D. Electromagnetic Wave Propagation in Media with Indefinite Permittivity and Permeability Tensors. *Phys. Rev. Lett.* **90,** 077405 (2003).

25. Siddiqui, O. & Eleftheriades, G. V. Resonance-cone focusing in a compensating bilayer of continuous hyperbolic microstrip grids. *Appl. Phys. Lett.* **85,** 1292 (2004).

26. Ishii, S., Kildishev, A. V., Narimanov, E., Shalaev, V. M. & Drachev, V. P. Sub-wavelength interference pattern from volume plasmon polaritons in a hyperbolic medium, *Laser Photonics Rev.* **7,** 265-271 (2013).

27. Thongrattanasiri, S. & Podolskiy, V. A. Hypergratings: nanophotonics in planar anisotropic metamaterials. *Opt. Lett.* **34,** 890-892 (2009).

28. Ishii, S., Drachev, V. P. & Kildishev, A. V. Diffractive nanoslit lenses for subwavelength focusing. *Opt. Commun.* **285**, 3368-3372 (2012).

29. Rho, J., Ye, Z., Xiong, Y., Yin, X., Liu, Z., Choi, H., Bartal, G. & Zhang, X. Spherical hyperlens for two-dimensional sub-diffractional imaging at visible frequencies. *Nature Commun.* **1,** 143 (2010).

30. Jacob, Z., Alekseyev, L. V. & Narimanov, E. Optical Hyperlens: Far-field imaging beyond the diffraction limit. *Opt. Express* **14,** 8247-8256 (2006).

31. Salandrino, A. & Engheta, N., Far-field subdiffraction optical microscopy using metamaterial crystals: Theory and simulations, *Phys. Rev. B* **74,** 075103 (2006).

32. Lu, D. & Liu, Z. Hyperlenses and metalenses for far-field super-resolution imaging. *Nature Commun.* **3,** 1205 (2012).

33. Dai, S. et al. Tunable phonon polaritons in atomically thin van der waals crystal of boron nitride. *Science* **343,** 1125-1129 (2014).

34. Caldwell, J. D. et al. Sub-diffraction, Volume-confined Polaritons in the Natural Hyperbolic Material, Hexagonal Boron Nitride, *Nature Commun*. **5**, 5221 (2014).

35. Geim, A. K., & Grigorieva, I. V. Van der Waals heterostructures. *Nature* **499,** 419-425 (2013).

36. Fogler, M. M., Butov, L. V. & Novoselov, K. S., High-temperature super-fluidity with indirect excitons in van der Waals heterostructures, *Nature Commun.* **5,** 4555 (2014).

37. Xu, X. G. et al. One-dimensional surface phonon polaritons in boron nitride nanotubes. *Nature Commun.* **5**, 4782 (2014).

38. Tarkhanyan, R. & Uzunoglu, N., Radiowaves and Polaritons in Anisotropic Media: Uniaxial Semiconductors (Wiley, Hoboken, 2006).

39. Novotny, L. & Hecht, B. Principles Of Nano-Optics (Cambridge University Press, Cambridge, 2006).

40. Atkin, J. M., Berweger, S., Jones, A. C. & Raschke, M. B. Nano-optical imaging and spectroscopy of order, phases, and domains in complex solids. *Adv. Phys.* **61,** 745-842 (2012).

41. Chaves, J. Introduction to Nonimaging Optics (CRC Press, Boca Raton, FL, 2008).

42. Chen, J. et al. Optical nano-imaging of gate-tunable graphene plasmons. *Nature* **487,** 77-81 (2012).

43. Fei, Z. et al. Gate-tuning of graphene plasmons revealed by infrared nano-imaging. *Nature* **487,** 82-85 (2012).

44. Woessner, A. et al. Highly confined low-loss plasmons in graphene-boron nitride heterostructures. *Nature Mater.* (2015).

45. Li, P. et al. Hyperbolic Phonon-polaritons in Boron Nitride for near-field optical imaging. *arXiv* (2015). http://arxiv.org/abs/1502.04093

46. Boltasseva, A., Atwater, H. Low-loss Plasmonic Metamaterials. *Science* **331,** 290-291 (2011).

47. Tassin, P., Koschny, T., Kafesaki, M. & Soukoulis, C. A comparison of graphene, superconductors and metals as conductors for metamaterials and plasmonics. *Nat. Photon.* **6,** 259-264 (2012).

48. Khurgin, J. B. & Boltasseva, A. Reflecting upon the losses in plasmonics and metamaterials. *MRS Bull.* **37**, 768-779 (2012).

49. Keilmann, F. & Amarie, S. Mid-infrared frequency comb spanning an octave based on an Er fiber laser and difference-frequency generation. *J. Infrared Millimeter Terahertz Waves* **33,** 479-484 (2012).

50. Watanabe, K., Taniguchi, T. & Kanda, H. Direct-bandgap properties and evidence for ultraviolet lasing of hexagonal boron nitride single crystal. *Nature Mater.* **3,** 404-409 (2004).

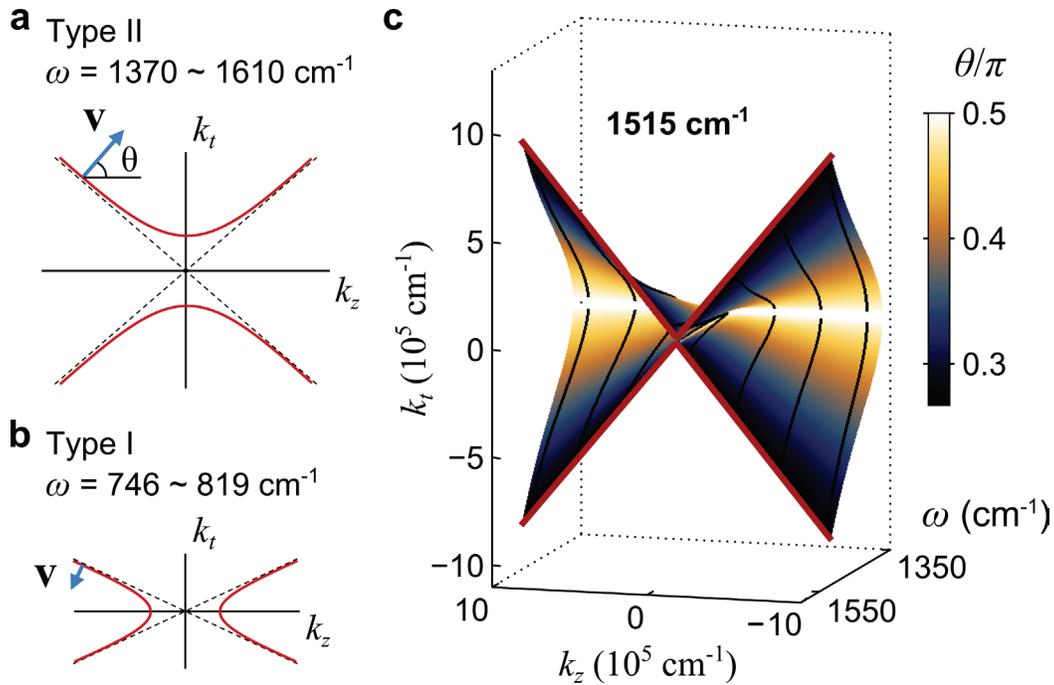

**Figure 1 | Hyperbolic dispersion of hBN. a**, A sketch of the isofrequency curves for a Type II HM, which is realized in the upper stop-band of hBN. The arrow indicates the polariton group velocity. **b**, A similar sketch for the Type I case, which is realized in the hBN lower stop-band. **c**, The calculated dispersion surface of hBN polaritons. The axes are the tangential momentum ($k_t$), the axial momentum ($k_z$), and the frequency ($\omega$, ranging from 1370 to 1515 cm$^{-1}$). The color represents the propagation angle $\theta$. The constant-frequency cut $\omega = 1515$ cm$^{-1}$ is shown by the red line, to emphasize similarity with (**a**). The dispersion of polaritons in a finite-thickness crystal ($d = 105$ nm) is shown by the black lines, to clarify their relation to fig. 4a.

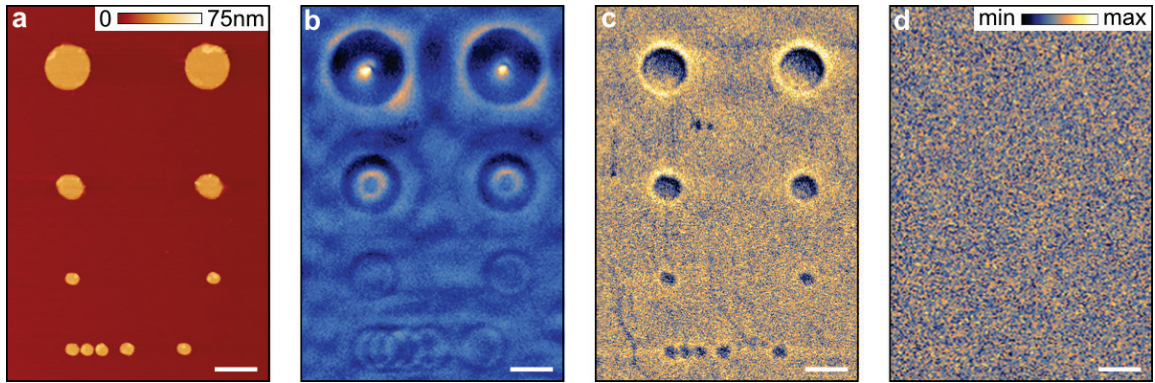

**Figure 2 | Sub-diffractional focusing and imaging through an hBN crystal. a**, An AFM image of Au disks defined lithographically on $SiO_2$/Si substrate before hBN transfer. **b**, Near-field amplitude image of the top surface of a 395-nm-thick hBN at IR laser frequency $\omega = 1515$ cm$^{-1}$ ($\lambda = 6.6$ μm). The observed "hot rings" are concentric with the Au discs. **c**, Near-field image of the same sample as in panel (**b**) at $\omega = 1610$ cm$^{-1}$ ($\lambda = 6.2$ μm) where polaritons propagate almost vertically. **d**, Near-field image of the same sample at $\omega = 1740$ cm$^{-1}$ ($\lambda = 5.7$ μm) showing complete homogeneity and lack of any distinct features. The color scales for (**b – d**) are indicated in (**d**). The scale bars in all panels are 1μm long.

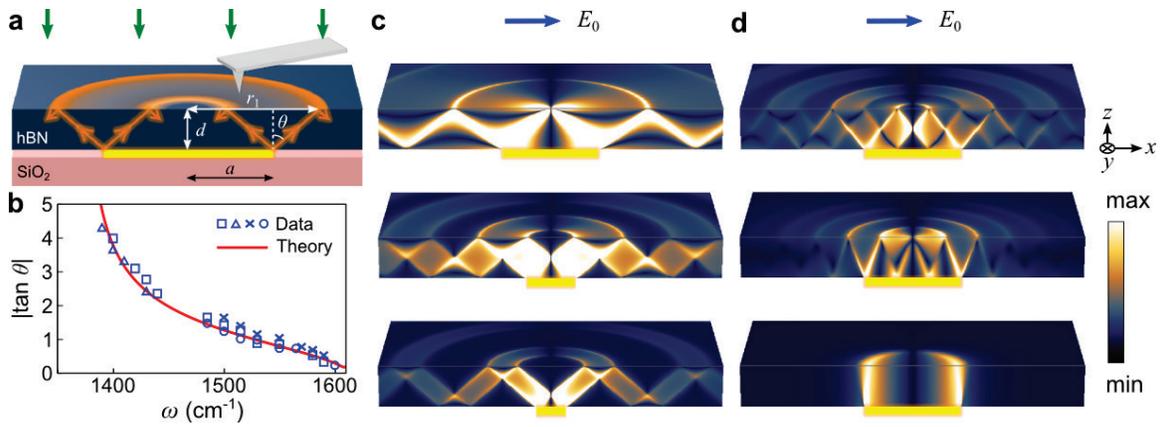

**Figure 3 | Image formation. a**, Imaging schematics. Under IR illumination (green arrow), the polaritons were launched by the Au disk edges and propagate towards the hBN top surface where the near-field images were recorded via the back-scattered IR beam (green arrow). The propagation angle $\theta$ can be inferred from the hot ring radius $r_1$, hBN thickness $d$, and disk radius $a$. **b**, The tangent of the propagation angle $\theta$ derived from imaging data for different hBN samples (symbols) and from equation (2) (solid line). Squares, triangles, crosses and dots indicate data from hBN samples with thickness $d$ = 395, 984, 270 and 1060 nm, respectively. **c**, The distribution of the $z$-component of the electric field in the analytical model (see text). The hot rings on the surfaces appear as a result of multiple reflections of polaritons launched at the disk edges. The ratio $a/|\delta|$ = 0.5, 0.25, 0.15 decreases from top to bottom. In the top picture the smallest ring shrinks to a focal point. The blue arrow indicates the direction of electric field $E_0$ in simulation. **d**, Similar to panel (**b**) for $a/d$ = 1.12 and (top to bottom) $|\tan\theta|$ = 0.75, 0.375 and 0.01.

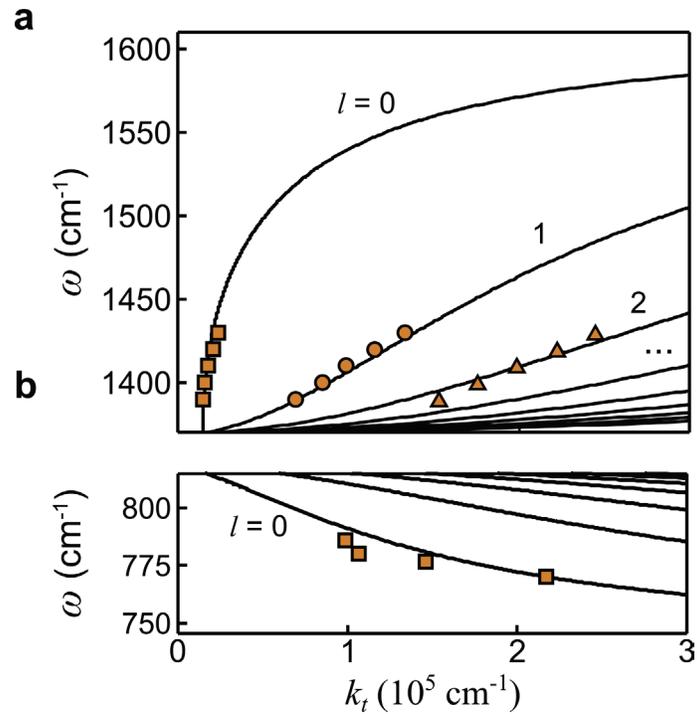

**Figure 4 | Polariton frequency ($\omega$) – in-plane momentum ($k_t$) dispersion relation for hBN. a**, The dispersion curves from Fig. 1c replotted as frequency ($\omega$) *vs*. in-plane momenta ($k_t$). The experimental data (squares) are obtained from the polariton reflection images near the sample edges (Fig. 5). **b**, Same as (**a**) for the lower hBN stop-band (Supplementary Figure 3). Thickness of hBN: 105 nm.

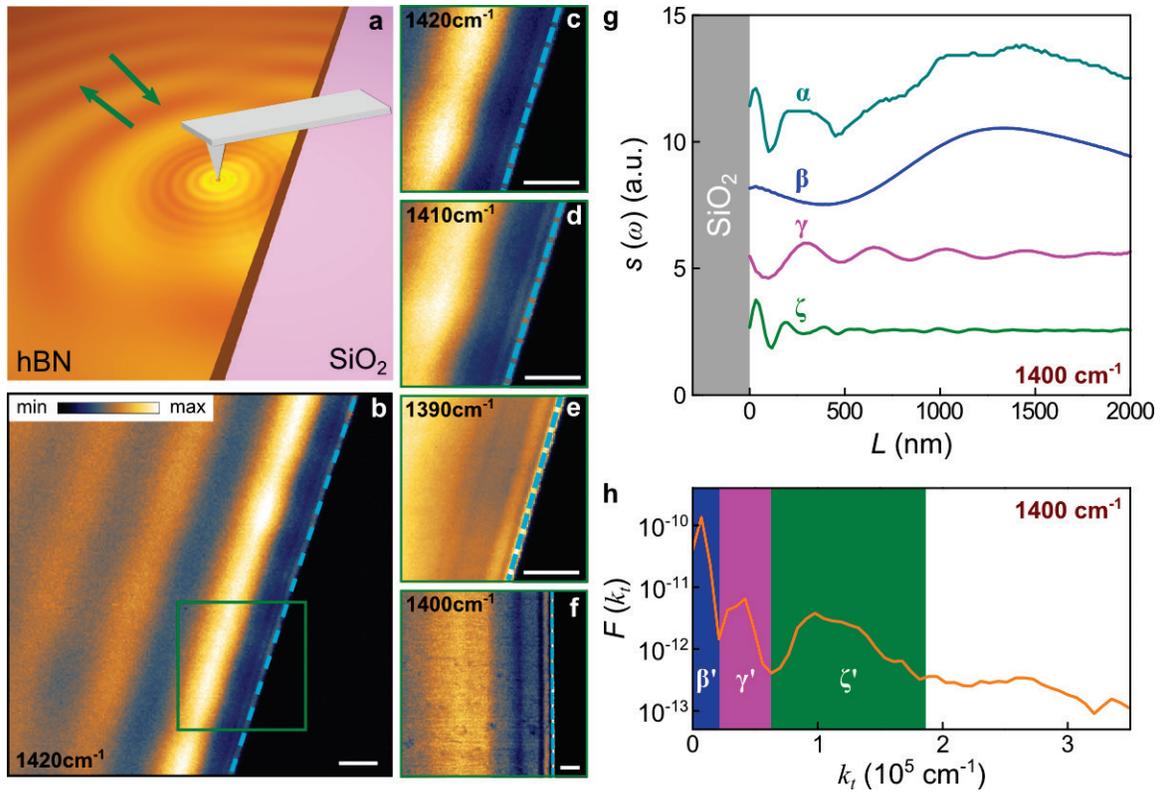

**Figure 5 | Imaging of polariton waveguide modes near the hBN edges. a**, Experimental schematic is similar to Fig. 3a except imaging here is performed near the edge of an unpatterned sample. **b,** Near-field amplitude image measured at 1420 cm$^{-1}$. The olive square indicates the area whose expanded view is shown in (**c-e**). **c-e**, Near-field image of the area marked in (**b**) at several frequencies. hBN thickness in (**b−e**): 31 nm. **f**, Near-field image of 105-nm-thick hBN at 1400 cm$^{-1}$. The cyan dashed lines in (**b−f**) indicate the hBN edges. Scale bar in (**b−f**): 300 nm. **g**, Line traces perpendicular to the hBN edge**.** Trace α was extracted from the image in panel **f**. Traces β, γ and ζ were obtained from the Fourier analysis of the trace α as described in the text. **h**: The Fourier transform of trace α in panel **g**.